\documentclass[aps,prl,twocolumn,amsmath,10pt,superscriptaddress]{revtex4-1} 
\pdfoutput=1
\usepackage{feynmp, tensor}

\usepackage{ulem}

\usepackage{color,graphicx}
\usepackage{bm}

\usepackage{amsmath}
\usepackage{amsfonts}
\usepackage{amssymb}

\usepackage{subfigure}
\usepackage[latin1]{inputenc}

\renewcommand{\-}{\,-\,}

\newcommand{\be}{\begin{equation}}
\newcommand{\ee}{\end{equation}}

\newcommand{\bea}{\begin{equation}\begin{aligned}}
\newcommand{\eea}{\end{aligned}\end{equation}}

\let\oldmarginpar\marginpar
\renewcommand\marginpar[1]{\-\oldmarginpar[\raggedleft\tiny #1]%
{\raggedright\tiny #1}}

\newcommand{\subfigimg}[3][,]{%
  \setbox1=\hbox{\includegraphics[#1]{#3}}
  \leavevmode\rlap{\usebox1}
  \rlap{\hspace*{-5pt}\raisebox{\dimexpr\ht1-2\baselineskip}{#2}}
  \phantom{\usebox1}
}
\begin{document}


\title{Hydrodynamic Electron Flow and Hall Viscosity}

\author{Thomas Scaffidi}
\affiliation{Department of Physics, University of California, Berkeley, CA 94720, USA}
\author{Nabhanila Nandi}
\affiliation{Max Planck Institute for Chemical Physics of Solids, Nöthnitzer Strasse 40, 01187 Dresden, Germany}
\author{Burkhard Schmidt}
\affiliation{Max Planck Institute for Chemical Physics of Solids, Nöthnitzer Strasse 40, 01187 Dresden, Germany}
\author{Andrew P. Mackenzie}
\affiliation{Max Planck Institute for Chemical Physics of Solids, Nöthnitzer Strasse 40, 01187 Dresden, Germany}
\affiliation{Scottish Universities Physics Alliance, School of Physics and Astronomy, University of St. Andrews, St. Andrews, Fife KY16 9SS, UK}
\author{Joel E. Moore}
\affiliation{Department of Physics, University of California, Berkeley, CA 94720, USA}
\affiliation{Materials Sciences Division, Lawrence Berkeley National Laboratory, Berkeley, CA 94720}

\date{\today}
\pacs{
}

\begin{abstract}
In metallic samples of small enough size and sufficiently strong momentum-conserving scattering, the viscosity of the electron gas can become the dominant process governing transport. In this regime, momentum is a long-lived quantity whose evolution is described by an emergent hydrodynamical theory.
Furthermore, breaking time-reversal symmetry leads to the appearance of an odd component to the viscosity called the Hall viscosity, which has attracted considerable attention recently due to its quantized nature in gapped systems but still eludes experimental confirmation. Based on microscopic calculations, we discuss how to measure the effects of both the even and odd components of the viscosity using hydrodynamic electronic transport in mesoscopic samples under applied magnetic fields.
\end{abstract}

\maketitle 

The semiclassical theory of electronic conduction, based on relaxation of total momentum by impurities, phonons and umklapp scattering, occupies a central place in condensed matter physics.
It is therefore of particular interest to study the cases for which it fails.
One case that has attracted much interest is the possibility of a hydrodynamic regime, where transport is dominated by viscous effects \cite{gurzhi1963minimum,gurzhi1968hydrodynamic,PhysRevB.21.3279,PhysRevLett.52.368,gurzhi1995electron,molenkamp1994observation,PhysRevB.49.5038,PhysRevB.51.13389,PhysRevLett.52.368,PhysRevLett.71.2465,PhysRevLett.77.1143,Spivak20062071,PhysRevLett.106.256804,PhysRevB.92.165433,PhysRevLett.117.166601,PhysRevLett.113.235901,levitov,PhysRevB.92.165433,PhysRevB.94.125427,2016arXiv160707269G,2016arXiv161209239G,2016arXiv161200856L,2016arXiv161209275L,2017arXiv170304522G}.
One needs a large separation of scales between momentum-relaxing and momentum-conserving scattering in order to see these effects.
This was recently achieved in graphene\cite{Bandurin1055,Crossno1058} and PdCoO$_2$\cite{Moll1061,0034-4885-80-3-032501}.

Interest in such a hydrodynamic regime also emanated from a conjectured bound on diffusion constants for the hydrodynamics of strongly interacting quantum systems \cite{PhysRevLett.94.111601,Hartnoll}.
Even though the physics described in this work is semiclassical and probably still quite far from these quantum-mechanical bounds, the observations that we hope to stimulate would constitute an important first step towards the understanding of emergent hydrodynamical regimes in electronic systems.

A further motivation for the work is that reaching a viscous regime for a charged fluid enables one to break time-reversal symmetry by adding a magnetic field and hence to study a non-dissipative component to the viscosity tensor called the Hall viscosity.
The recent interest in this Hall viscosity emanates from the fact that it is topologically quantized in gapped systems \cite{PhysRevB.79.045308}.
In order to study this effect experimentally, in analogy with the Hall conductivity, the first step would obviously be to measure the {\it classical} Hall viscosity.
We show in this letter how this measurement could be done by describing specific size effects from Hall viscosity in transport in restricted 2D channels under transverse magnetic fields.

%
%
%
%

This paper is organized as follows. 
We start by assuming a perfect hydrodynamic regime and calculate $\rho_{xx}$ and $\rho_{xy}$.
We show that the $1/W^2$ component of $\rho_{xy}$ is proportional to the Hall viscosity, thereby providing a way of measuring it.
In order to have realistic predictions to compare with experiments, one should also take into account other, non-viscous effects that can lead to a size-dependent resistivity.
We thus perform a kinetic Boltzmann calculation in which the effect of diffuse boundaries, gradient along the section of the wire, momentum-conserving scattering, and magnetic field are taken into account.
We show that the size effects in resistivities with and without momentum-conserving scattering are markedly different, thereby making it possible to distinguish hydrodynamic and non-hydrodynamic size effects.
Finally, we comment on how these measurements could also enable one to measure the quantum Hall viscosity in the quantum Hall regime and establish a relation similar to Hoyos-Son \cite{PhysRevLett.108.066805}.

\section{Fluid equation}

Even though a finite amount of momentum relaxation is always present, it is instructive to first look at the limit where it is zero.
In this limit, the momentum density of the electron gas is conserved and one can write a hydrodynamic equation to model its dynamics \cite{PhysRevLett.117.166601} \footnote{Following Ref. \cite{PhysRevLett.117.166601}, we consider the case of an isochoric flow, $\nabla \cdot \vec{v} = 0$. Even though our solution exhibits non-zero variations in the local electron density compared to the average density, the pressure contribution coming from those variations is negligible.}:
\bea
\partial_t \vec{v} = \eta_{xx} \nabla^2 \vec{v} + \eta_{xy}  \nabla^2\vec{v} \times \vec{z}  + \frac{e}{m} (\vec{E} + \vec{v} \times \vec{B})
\eea
where $m$ is the electron mass and $\eta_{xx}$ and $\eta_{xy}$ are the regular and Hall components of the kinematic viscosity tensor.
As mentioned previously, the two diffusion constants in this equation are interesting from a fundamental point of view: (1) from a holographic argument, a bound on $\eta_{xx}$ was conjectured (coming from the bound on the dissipative time scale $\hbar/k_B T$) and (2) $\eta_{xy}$ is non-dissipative (and therefore not subject to this bound) but was shown to be quantized in gapped systems \cite{PhysRevLett.75.697,PhysRevB.84.085316,PhysRevD.88.025040,doi:10.1142/S0217979214300072}.

We consider the case of a two-dimensional channel of width $W$ along the $y$ direction, with a uniform applied electric field $E_x$ along $x$, uniform magnetic field $B$ along $z$, and zero current along $y$. In the stationary regime, one finds
\bea
\eta_{xx} \frac{d^2 v_x}{dy^2} + \frac{e}{m} E_x  &= 0 \\
-\eta_{xy} \frac{d^2 v_x}{dy^2}  + \frac{e}{m} E_y &=  \omega_c v_x 
\eea
where $\omega_c=eB/m$.
Using the conventional no-slip boundary condition $v_x(y=\pm W/2)=0$ (we will treat the more realistic case of diffuse boundaries within a kinetic formalism in the next section), one finds
\bea
\rho_{xx}  &=  \frac{m}{e^2 n} \eta_{xx} \frac{12}{ W^2} \\
\rho_{xy} &=   \rho_{xy}^{bulk} \left(1- \eta_{xy} \frac{12}{ W^2} \frac1{\omega_c} \right) 
\label{Resist}
\eea
where  $\rho_{xy}^{bulk}=-\frac{m \omega_c}{e^2 n}$. 
In the wide sample limit ($W \rightarrow \infty$), one has $\rho_{xx}=0$, since the only source of resistance comes from the boundaries, and $\rho_{xy} = -m\omega_c/e^2 n$, which is indeed its bulk value.
We propose to measure $\eta_{xx}$ and $\eta_{xy}$ by measuring the size dependence of $\rho_{xx}$ and $\rho_{xy}$ in restricted channels of varying size and under varying magnetic fields.

In the next section, we will compare these results with the results of a kinetic Boltzman formalism.
In order to do so, it is convenient to inject a microscopically derived magnetic field dependence of the viscosities in the above hydrodynamic solutions.
The following dependence was found by Alekseev \cite{PhysRevLett.117.166601}: 
\bea
\eta_{xx} &= \eta \frac{1}{1 + (2 \frac{l_{MC}}{r_c})^2} \\
\eta_{xy} &= \eta \frac{ 2 \frac{l_{MC}}{r_c}}{1 + (2 \frac{l_{MC}}{r_c})^2}
\label{Viscosities}
\eea
where $r_c=mv_F/eB=v_F/\omega_c$ is the cyclotron radius, $l_{MC} = v_F \tau_{MC}$ is the momentum-conserving scattering length and where $\eta = \frac14 v_F l_{MC}$  \cite{PhysRevLett.117.166601}.
This leads to
\bea
\rho_{xx} &=  \frac{m}{e^2 n} \frac{12}{ W^2}\eta \frac{1}{1 + (2 \frac{l_{MC}}{r_c})^2} \\
\rho_{xy} &= \rho_{xy}^{\text{bulk}}\left( 1 - 6  \frac{ 1 }{1 + (2 \frac{l_{MC}}{r_c})^2} \left( \frac{l_{MC}}{ W} \right)^2 \right)
\label{HydroPredictions}
\eea

The size effect on $\rho_{xy}$ is maximal at zero field due to the Lorentzian factor.
In this limit, one obtains
\bea
\rho_{xy}  &= \rho_{xy}^{\text{bulk}} \left( 1 - 6   \left( \frac{l_{MC}}{ W} \right)^2 \right) \text{ for } B \rightarrow 0
\eea
In order to measure this effect, $W$ should be as small as possible for it to be sizable, but still somewhat larger than $l_{MC}$ in order to remain in the hydrodynamic regime.
For example, for $W/l_{MC}=5 $, one expects a relative change of the Hall slope at zero field of the order of $25 \%$, which should be measurable.


%
%
%

\section{Kinetic theory}
As mentioned previously, in order to make a quantitative comparison with experiments, it is crucial to go beyond a purely hydrodynamical theory and take into account several other effects like the non-zero momentum relaxation and the diffuse scattering at the boundaries.
In order to do this, we perform kinetic Boltzmann calculations \cite{doi:10.1080/14786436608211970,PhysRevB.51.13389,Moll1061}:
\bea
\vec{v} \cdot \nabla_{\vec{r}} f + \frac{e}{m} (E + \vec{v} \times \vec{B}) \cdot \nabla_{\vec{v}} f = \left. \frac{\partial f}{\partial t} \right|_{\text{scatt}}
\label{Boltzman}
\eea
where $\vec{B}$ is the external magnetic field, $f(\vec{r},\vec{v})$ is the semiclassical occupation number for a wavepacket at position $\vec{r}$ and velocity $\vec{v}$ and where
\bea
\left. \frac{\partial f(\vec{r},\vec{v})}{\partial t} \right|_{\text{scatt}} = -\frac{f(\vec{r},\vec{v}) - n(\vec{r})}{\tau} + \frac2{\tau_{MC}} \vec{v} \cdot \vec{j}(\vec{r})
\eea
with $n(\vec{r})=\left\langle f \right\rangle_{\vec{v}}$ the local charge density, $\vec{j}(\vec{r})=\left\langle f \vec{v}\right\rangle_{\vec{v}}$ the local current, $\left\langle \dots  \right\rangle_{\vec{v}}$ the momentum average and $\tau^{-1}=\tau_{MR}^{-1} + \tau_{MC}^{-1}$.
For the sake of simplicity, we consider the case of a circular Fermi surface with $\vec{v} = v_F \hat{\rho}$ with $\hat{\rho}$ the radial unit vector.
Scattering lengths are then simply defined as $l_{MR(MC)}=v_F \ \tau_{MR(MC)}$.
The term proportional to $\tau_{MC}^{-1}$ is the most simple momentum-conserving scattering term that can be written assuming that the electrons relax to a Fermi-Dirac distribution shifted by the drift velocity \cite{PhysRevB.49.5038,Moll1061}.
%
%
The boundary conditions are given by
\bea
j_y(y=\pm W/2)&=0 \\
f(y=\pm W/2,\vec{v}) &= \pm f_{\text{boundary}} 
\eea
which imposes, respectively, no current in the $y$ direction and diffuse scattering at the boundaries.
Equation \ref{Boltzman} is supplemented by Gauss's law with a charge density given by $e n(\vec{x})$.
The resulting integrodifferential equation is advantageously solved numerically by using the method of characteristics \cite{Charac}.

Three limiting regimes can be identified \cite{PhysRevB.51.13389}: Ohmic, hydrodynamic and ballistic.
In the Ohmic case, one has $l_{MR} \ll W$, and transport is therefore dominated by momentum-relaxing scattering, leading to bulk values for transport coefficients: $\rho_{xx}=\rho_{xx}^{\text{bulk}}=m/n e^2 \tau_{MR}$ and $\rho_{xy}=\rho_{xy}^{\text{bulk}}=-m \omega_c / e^2 n$.
In the ballistic case, one has $W \ll l_{MR}, l_{MC}$, and transport is dominated by scattering with the boundaries. This leads to strong size effects, but with a different size dependence from the one arising in the hydrodynamic regime.
Finally, in the hydrodynamic case, one has $l_{MC} \ll W \ll l_{MR}$, and transport is dominated by the diffusion of momentum at the boundaries through viscosity.
Note that this regime will only appear if $l_{MC} \ll l_{MR}$.

\begin{figure}
  \centering
  \begin{tabular}{@{}p{0.75\linewidth}}
    \subfigimg[width=\linewidth]{a)}{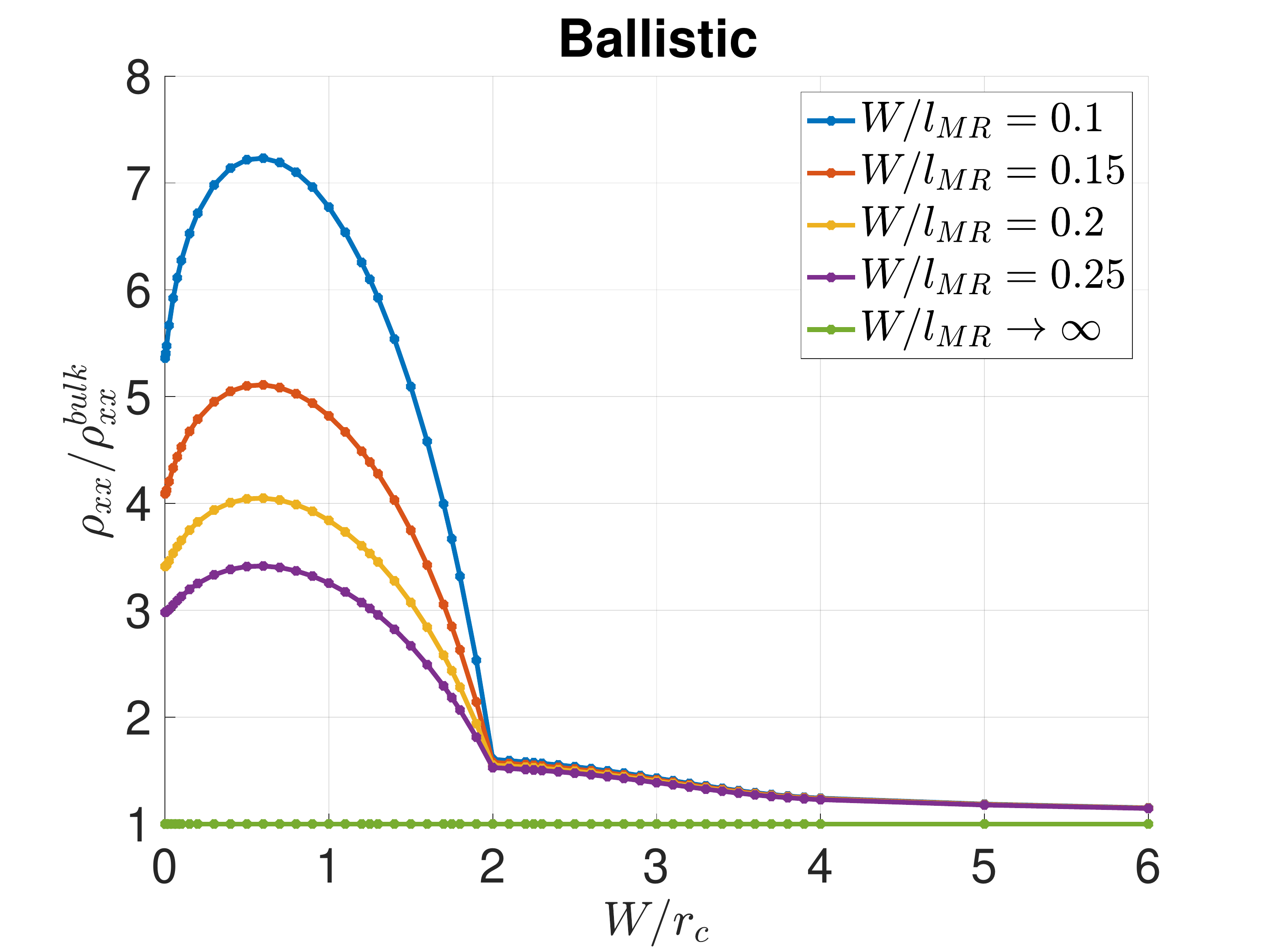}  \\
    \subfigimg[width=\linewidth]{b)}{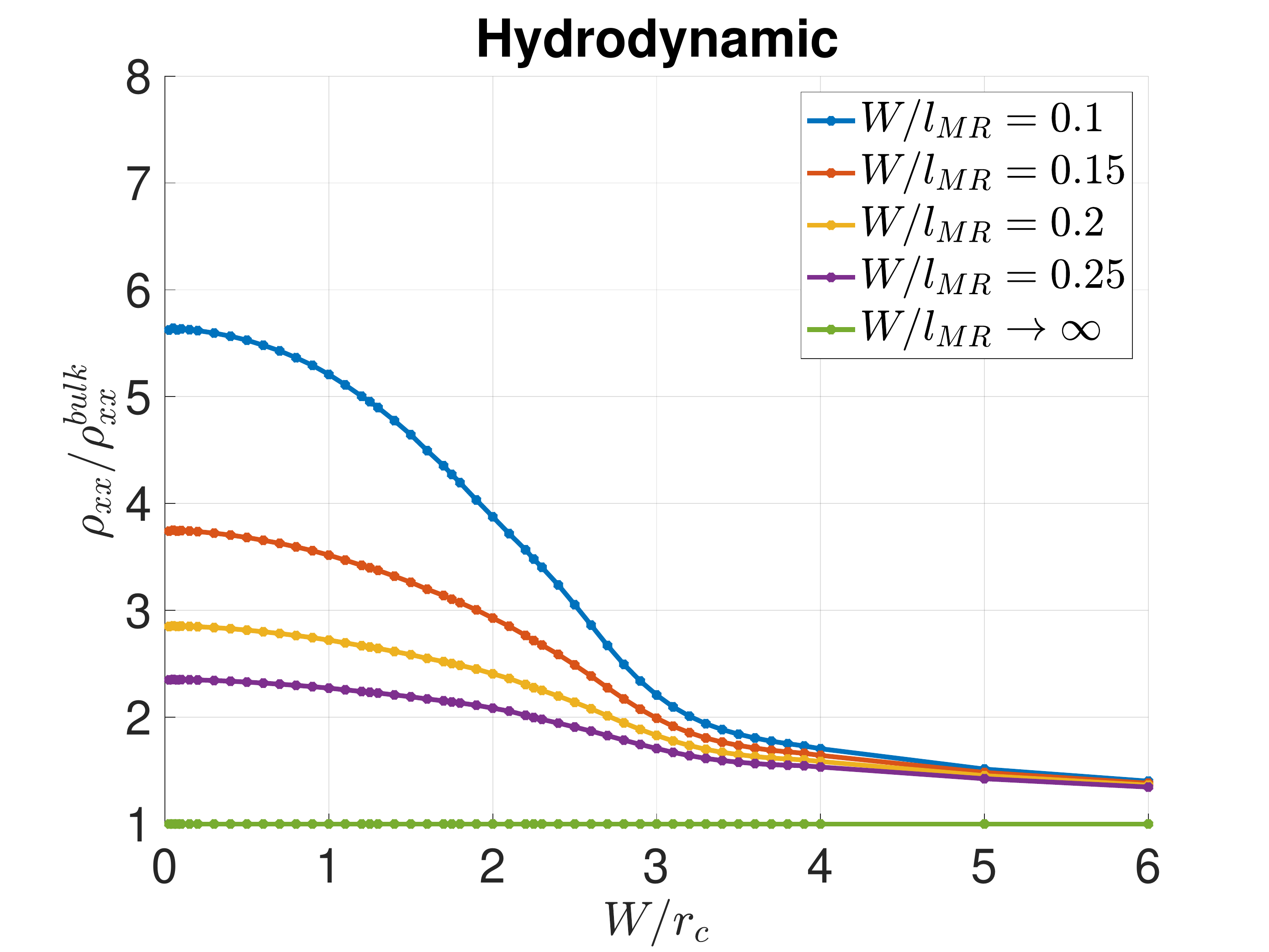} 
  \end{tabular}
\caption{Magnetoresistance for the ballistic ($l_{MC}/l_{MR}=10$) (a) and hydrodynamic case ($l_{MC}/l_{MR}=0.05$) (b).
\label{RhoXX}}
\end{figure}

\begin{figure}
  \centering
  \begin{tabular}{@{}p{0.75\linewidth}}
    \subfigimg[width=\linewidth]{a)}{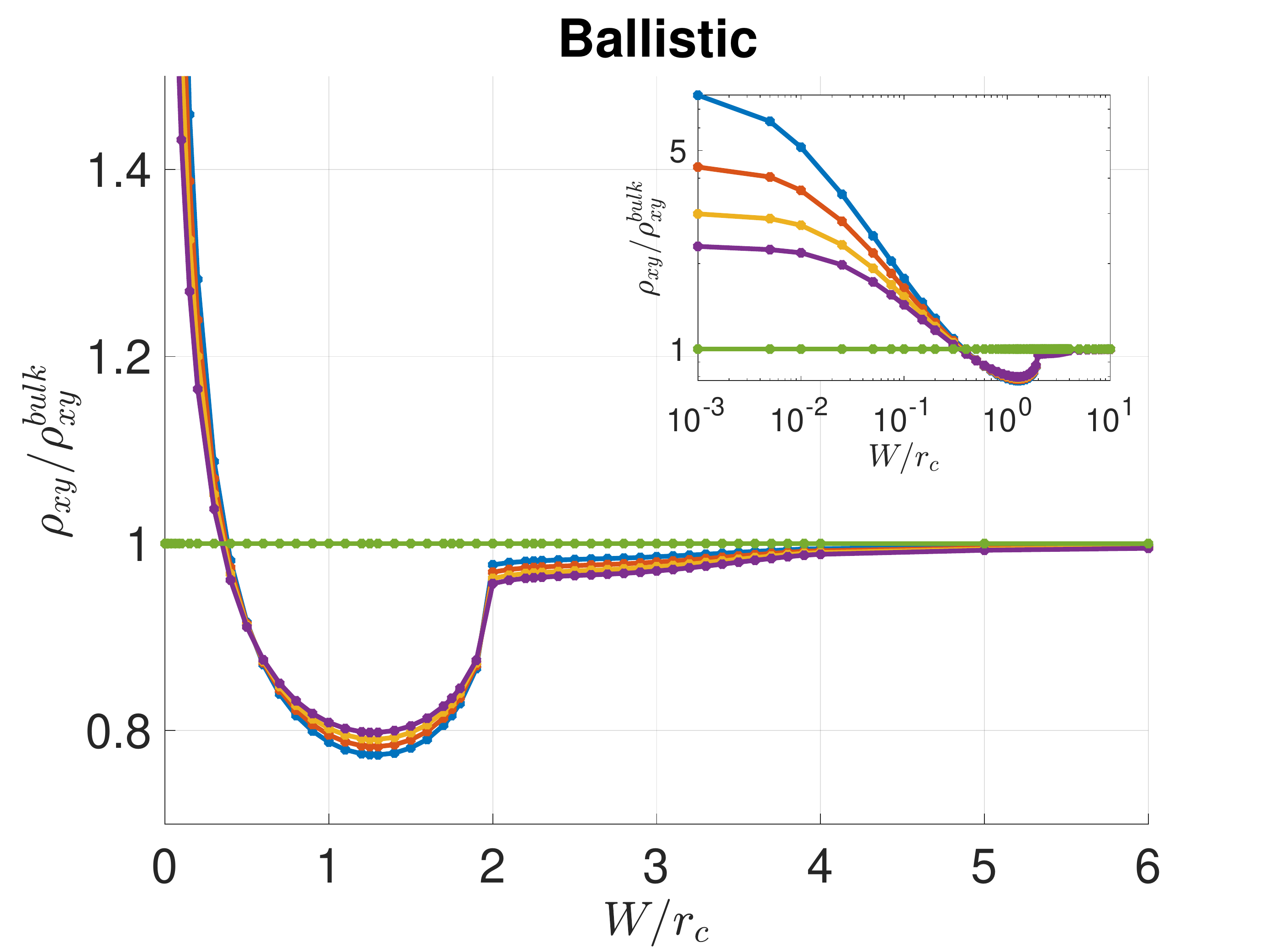}  \\
    \subfigimg[width=\linewidth]{b)}{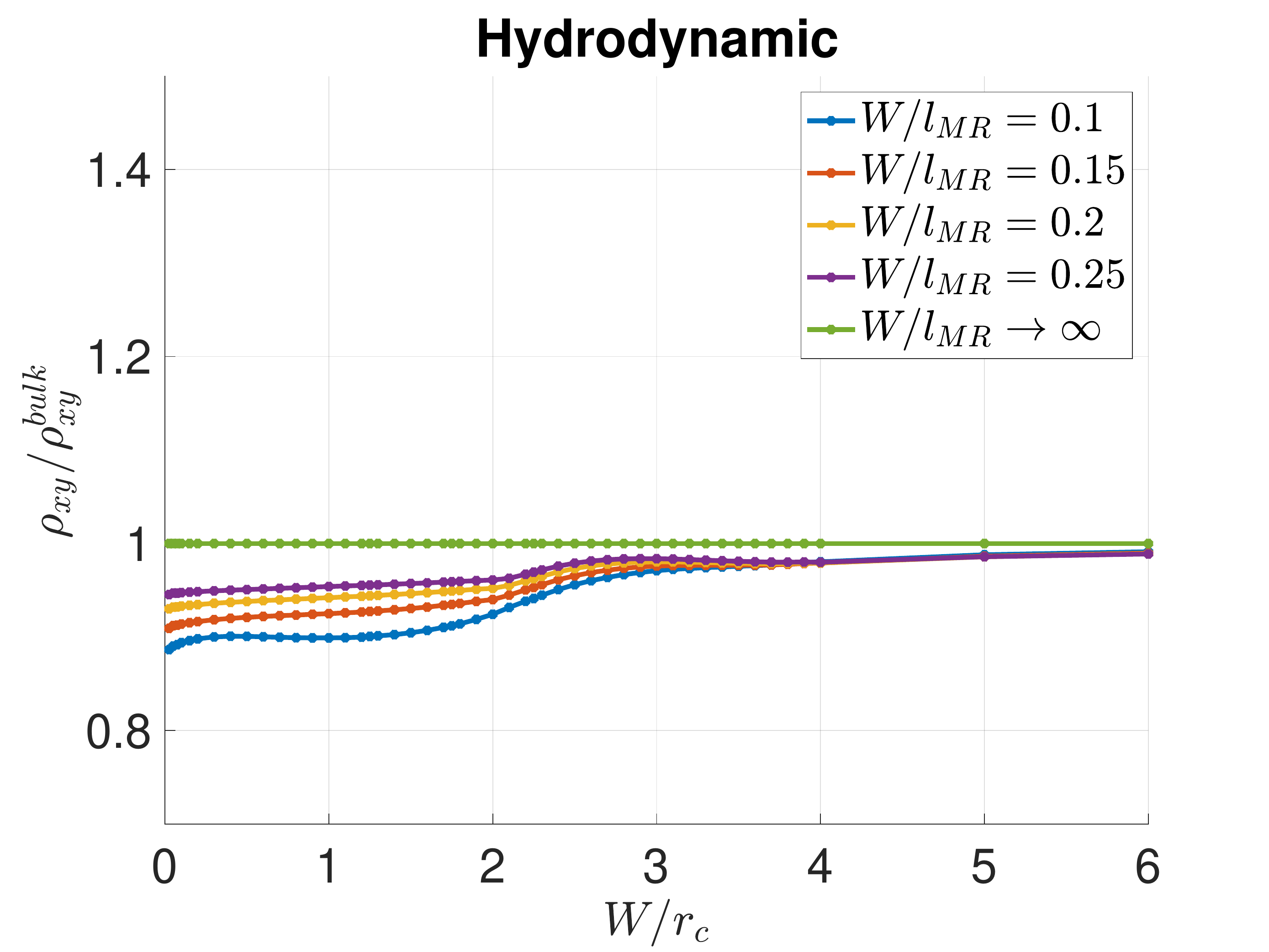} 
  \end{tabular}
\caption{Hall resistivity for the ballistic ($l_{MC}/l_{MR}=10$) (a) and hydrodynamic case ($l_{MC}/l_{MR}=0.05$) (b). The inset of panel (a) shows the saturation of the Hall resistivity at low fields in the ballistic case.
\label{RhoXY}}
\end{figure}

Using the above kinetic theory, we look at transport in and between these different regimes and identify clear signatures of a hydrodynamic regime.
In figures \ref{RhoXX} and \ref{RhoXY}, we give respectively the magnetoresistance and the Hall resistivity in a typical ballistic case ($l_{MC}/l_{MR}=10$, $W \lesssim l_{MR}$) and a typical hydrodynamic case ($l_{MC}/l_{MR}=0.05 $, $l_{MC} < W < l_{MR}$).

In the ballistic regime (Fig 1.a), the magnetoresistance shows a maximum around $W/r_c \simeq 0.55$ and a rapid change of slope at $W/r_c=2$, as previously reported \cite{doi:10.1080/14786436608211970,Beenakker19911}.
In this ballistic regime, transport is dominated by electrons with a velocity close to the longitudinal ($\hat{x}$) direction for which scattering on boundaries are very rare.
These trajectories are bent by the field, leading to an increase in boundary scattering and therefore an increase of $\rho_{xx}$ at low fields.
At higher fields, when the cyclotron radius becomes of the order of $W$, a larger and larger fraction of electrons present near the middle of the wire stop seeing the boundaries at all, and at large fields the bulk resistivity is therefore recovered.

In contrast, in the hydrodynamic regime (Fig 1.b), the longitudinal magnetoresistance decays monotonically and there is no sharp slope change at $W/r_c=2$.
The magnetoresistance follows a Lorentzian-like curve, in agreement with the hydrodynamic calculation of Eq. \ref{HydroPredictions}.
This absence of sharp behavior at $W/r_c=2$ shows that, in this case, momentum-conserving scattering is dominant and no electron can ever go over a full cyclotron orbit without being scattered.
As explicit in Eq. \ref{HydroPredictions}, the relevant dimensionless parameter is therefore now $l_{MC}/r_c$, as opposed to $W/r_c$ in the ballistic case.
The absence of a maximum of the magnetoresistance and of a sharp slope change at $W/r_c = 2$ can be used as clear signatures of the hydrodynamic regime.
Note that the different scaling of $\rho_{xx}$ with $W$ in zero applied field can also be used to differentiate the two regimes, as reported in \cite{Moll1061}.

The Hall resistivity also shows strikingly different behaviors in the ballistic and hydrodynamic regimes.
As seen in Fig. 2.b, in the ballistic case, $\rho_{xy}$ exhibits a minimum at $W/r_c \simeq 1.3$ and a sharp slope change at $W/r_c = 2$, in analogy with $\rho_{xx}$ and in agreement with the perturbative calculation given in Ref. \cite{PhysRev.80.401}.
While, in the ballistic case, $\Delta\rho_{xy}\equiv \rho_{xy}-\rho_{xy}^{\text{bulk}}$ is positive at low fields and negative at large fields, it is always negative in the hydrodynamic case (see Fig 2.b).
The sign of $\Delta\rho_{xy}$ can therefore be used a clear signature of hydrodynamic effects: one would expect a smaller (resp. larger) Hall slope at small fields than at large fields in the hydrodynamic (resp. ballistic) regime.

It was checked that, in both the cases of $l_{MC}/l_{MR}=10$ and $l_{MC}/l_{MR}=0.05$, if $W$ becomes larger than $l_{MR}$, Ohmic behavior is recovered.
It was also checked that, in the case of $l_{MC}/l_{MR}=0.05$, if $W$ becomes smaller than $l_{MR}$, ballistic behavior is recovered.


%

\section{Quantum limit}
While the calculations up to this point have been classical, it is instructive to look at the large magnetic field limit, $r_c/l_{MC} \rightarrow 0$.
In this limit, one finds from Eq. \ref{Viscosities} that
\bea
\eta_{xy} &\rightarrow \frac18 v_F r_c
\eea
which, crucially, does not depend on $l_{MC}$.
Now, using $\nu = 2\pi l_B^2 n$, $v_F = \hbar k_F / m$ with $k_F = \sqrt{4 \pi n}$ and $l_B^2 = \hbar/eB$, one finds
\bea
\tilde{\eta}_{xy} = n m \eta_{xy} = \frac1{8\pi} \frac{\hbar \nu^2}{l_B^2}
\eea
which gives the quantized value of the dynamical viscosity $\tilde{\eta}$ for a quantum Hall system at filling $\nu$.
Inverting the resistivity tensor obtained in Eq. \ref{Resist} and taking the large field limit leads to the conductivity:
\bea
\sigma_{xy} &=\nu\frac{e^2}{h}  \left(1 +   \frac{\tilde{\eta}_{xy}}{\hbar n}  (q_{\text{eff}} l_B)^2\right) \\
2\pi/q_{\text{eff}} &= 2\pi W/\sqrt{12} \simeq 1.8 W
\label{HoyosSonFromClassical}
\eea
The large magnetic field extrapolation of the classical calculation found in this paper is therefore consistent with the form of the finite-$q$ correction found by Hoyos and Son in the quantum case \cite{PhysRevLett.108.066805} \footnote{Our classical calculation deals with compressible fluids and is therefore not a priori expected to apply to incompressible quantum Hall fluids. Yet, out compressible calculation could still apply to a quantum Hall system with $W \lesssim l_B$, which is effectively a gapless system because of finite-size effects. Interestingly, this is the regime in which our calculations predict a sizable viscous contribution to $\sigma^{xy}$.}.
In the quantum Hall regime, the relative deviation of $\sigma_{xy}$ from its bulk value is of the order of $\nu^2 (q_{\text{eff}} l_B)^2$. 
For realistic system sizes, the factor $(q_{\text{eff}} l_B)^2$ would probably be too small to lead to a measurable deviation. 
One could then turn to optical probes where $q_{\text{eff}}$ could be chosen in the X-ray range.

\section{Conclusion}
In conclusion, we have identified clear, qualitative signatures of hydrodynamic behavior in transport measurements of mesoscopic metallic samples under magnetic fields.
This type of measurement is possible with readily available experimental techniques, and would make it possible to measure the classical Hall viscosity of the electron gas, which to the best of our knowledge has never been measured in a solid-state system.
This would both further the evidence of a hydrodynamic regime in electronic transport and constitute an important step towards an experimental understanding of the quantum Hall viscosity.

\section{Acknowledgements}
Helpful conversations with Tankut Can and Sriram Ganeshan are acknowledged.
The authors acknowledge support from the Emergent Phenomena in Quantum Systems initiative of the Gordon and Betty Moore Foundation (T. S.) and NSF DMR-1507141 and a Simons Investigatorship (J.E.M.).
We also acknowledge the support of the Max Planck Society and the UK Engineering and Physical Sciences Research Council under grant EP/I032487/1.
\bibliography{HydroPaper}

\clearpage 

\appendix
\end{document}